\documentclass[prb,twocolumn,reprint,nofootinbib,superscriptaddress,amsmath,amssymb,aps]{revtex4-1}

\pdfoutput=1

\usepackage{graphicx}
\usepackage{dcolumn}
\usepackage{bbm}
\usepackage{bm}
\usepackage[dvipsnames,x11names]{xcolor}
\usepackage{mathtools}
\usepackage{mathrsfs}
\usepackage{slashed}
\usepackage[protrusion=true,expansion,kerning=true,tracking=true,final]{microtype}
\usepackage{soul}
\usepackage{tabularx}
\usepackage[colorlinks=true,
	        linkcolor=hblue,
	        citecolor=hgreen,
	        filecolor=hblue,
	        urlcolor=hred
	        ]{hyperref}

\definecolor{hgreen}{rgb}{0,.3,0}
\definecolor{hred}{rgb}{.3,0,0}
\definecolor{orange}{rgb}{1,0.5,0}
\definecolor{hblue}{rgb}{0,0,.3}
\definecolor{LightGray}{gray}{0.95}
\definecolor{gray}{gray}{0.6}

\DeclareOldFontCommand{\rm}{\normalfont\rmfamily}{\mathrm}
\DeclareOldFontCommand{\sf}{\normalfont\sffamily}{\mathsf}
\DeclareOldFontCommand{\tt}{\normalfont\ttfamily}{\mathtt}
\DeclareOldFontCommand{\bf}{\normalfont\bfseries}{\mathbf}
\DeclareOldFontCommand{\it}{\normalfont\itshape}{\mathit}
\DeclareOldFontCommand{\sl}{\normalfont\slshape}{\@nomath\sl}
\DeclareOldFontCommand{\sc}{\normalfont\scshape}{\@nomath\sc}

\newcommand{\Lag}{\mathscr{L}}
\newcommand{\SO}{\mathrm{SO}}
\newcommand{\SU}{\mathrm{SU}}

\begin{document}

\title{Absence of SO$\mathbf{(4)}$ quantum criticality in Dirac semimetals at two-loop order}
\date{\today}

\begin{flushright}
      \normalsize DO-TH 23/13
\end{flushright}
\vspace{-0.8cm}

\author{Max~Uetrecht}
\affiliation{Fakult\"at Physik, Technische Universit\"at Dortmund, D-44221 Dortmund, Germany}

\author{Igor F.~Herbut}
\affiliation{Department of Physics, Simon Fraser University, Burnaby, British Columbia, Canada V5A 1S6}

\author{Emmanuel Stamou}
\affiliation{Fakult\"at Physik, Technische Universit\"at Dortmund, D-44221 Dortmund, Germany}

\author{Michael M.~Scherer}
\affiliation{Institut f\"ur Theoretische Physik III, Ruhr-Universit\"at Bochum, D-44801 Bochum, Germany}

\begin{abstract}
Evidence for relativistic quantum criticality of antiferromagnetism and
superconductivity in two-dimensional Dirac fermion systems has been found in
large-scale quantum Monte Carlo simulations.  However, the corresponding
($2+1$)-dimensional Gross--Neveu--Yukawa field theory with $N_f=2$ four-component
Dirac fermions coupled to two triplets of order parameters does not exhibit a
renormalization group fixed point at one-loop order.  Instead, the theory only
features a critical point for a large or very small fractional number of fermion flavors
$N_f$, which disappears for a broad range of flavor numbers around the physical
case, $N_f=2$, due to fixed-point annihilation.  This raises the question on
how to explain the observed scaling collapse in the quantum Monte Carlo data.
Here, we extend previous renormalization-group analyses by studying a
generalized model at two-loop order in $4-\epsilon$ spacetime dimensions.  We
determine the $\epsilon$ correction to the upper and lower critical flavor
numbers for the fixed-point annihilation and find that they both go towards the
physical case $N_f=2$.  However, this only happens very slowly, such that an
extrapolation to $\epsilon=1$ still suggests the absence of criticality in
$2+1$ dimensions.  Thereby, we consolidate the finding that the continuum field
theory does not feature a stable renormalization-group fixed point and no true
quantum criticality would be expected for the considered system. We briefly
discuss a possible reconciliation in terms of a complex conformal field theory.
Further, we also explore the fixed-point structure in an enlarged theory space
and identify a candidate stable fixed-point solution.
\end{abstract}

\maketitle

\section{Introduction\label{sec:introduction}}

The quantum critical behavior of two-dimensional gapless Dirac electrons undergoing a transition into an ordered phase is characterized by the generalized Gross--Neveu universality classes,\cite{Herbut2006,Herbut2009,Boyack:2020xpe,herbut2023wilson} which are distinct from the conventional universality classes known from statistical models.
The advent of Dirac materials,\cite{Vafek2014,Wehling2014} has boosted the interest in these universality classes~\cite{PhysRevB.83.035125,PhysRevB.87.041401,PhysRevB.89.205403,PhysRevB.94.245102,PhysRevB.94.205136,PhysRevB.93.155157,li2017fermion,PhysRevB.96.115132,PhysRevB.98.245120,PhysRevD.97.105009,PhysRevLett.120.215702,PhysRevB.103.075147,Yerzhakov:2020ote,PhysRevB.103.155160} which may even come within experimental reach, when electronic interactions are strong enough to induce symmetry-breaking phase transitions as, e.g., observed in magic-angle graphene at charge neutrality.\cite{lu2019superconductors,stepanov2020untying,PhysRevLett.123.157601,parthenios2023twisted}

The critical exponents of the Gross--Neveu universality classes depend on the
symmetries of the order-parameter field and the number of Dirac fermions.
For example, for the case of the chiral Ising model, their
precise computation has recently been pushed forward through progress in the
conformal bootstrap,\cite{Iliesiu:2015qra,Iliesiu:2017nrv,erramilli2023gross},  quantum Monte Carlo simulations~\cite{PhysRevD.88.021701,PhysRevB.101.064308,Wang:2023mjj} and renormalization
group methods,\cite{PhysRevB.98.125109,PhysRevB.96.165133,PhysRevD.96.096010} and a satisfactory agreement between the estimates for the exponents
across different theoretical approaches has already been achieved.

For other cases, however, whereas the different theoretical methods confirm the
existence of a distinct Gross--Neveu quantum critical point between the Dirac-semimetal phase and an
ordered state, the quantitative agreement between them remains more elusive. An important example of these is the chiral Heisenberg universality class that describes the transition towards an antiferromagnetic state in graphene,\cite{PhysRevD.96.096010,PhysRevX.3.031010,PhysRevB.91.165108,PhysRevX.6.011029,PhysRevB.102.245105,PhysRevB.102.235105} see the recent discussion presented in Ref.~[\onlinecite{herbut2023wilson}].

More work is required to achieve consensus on the precise value
of the critical exponents. Despite such quantitative differences,
these findings support the general picture that the critical behavior
near a continuous (quantum) phase transition is not governed by the
specifics of a microscopic model but that it can be captured in terms of a
universal continuum field theory.
This universal theory shares the symmetries, dimensionality, and
field content of the microscopic model as, e.g., required for
a quantum Monte Carlo simulation.

Interestingly, a recent quantum Monte Carlo study explored the transition of a system with two four-component Dirac fermions going from a semimetallic into a massive phase including a N\'eel state and a superconductor-CDW state.\cite{PhysRevLett.128.117202}
The quantum Monte Carlo data reported in Ref.~[\onlinecite{PhysRevLett.128.117202}] supports a scaling collapse indicative of a quantum critical point in the $\SO(3)\times \SO(3) \simeq \SO(4)$ -- symmetric field theory. In contrast, a subsequent renormalization-group study of the corresponding
continuum field theory could not identify a stable fixed point of the
model,\cite{Herbut:2022zzw} suggesting the absence of critical behavior
without additional fine-tuning, at least on the one-loop level studied in
Ref.~[\onlinecite{Herbut:2022zzw}].

These incompatible findings call for reconciliation and  options
for a resolution are the following:~{\itshape i)}~The lattice model studied
in~Ref.~[\onlinecite{PhysRevLett.128.117202}] corresponds to a different continuum field
theory, e.g., one with additional topological terms compatible with the symmetries.
{\itshape ii)}~The one-loop renormalization group misses the underlying fixed point, which could
be recovered at higher-loop orders.
{\itshape iii)}~The transition observed in the lattice model is of weak first
order merely mimicking critical behavior.

All of these options hint towards an interesting underlying physical mechanism,
e.g., the question about possible topological contributions according to~{\itshape i)} or
the similarity to the deconfined quantum phase transitions~\cite{senthilQuantum2004,sandvikEvidence2007,nahumDeconfined2015,senthilDeconfined2023} of option~{\itshape iii)}.
Option~{\itshape ii)} would have a certain similarity to the Banks--Zaks fixed point~\cite{Belavin:1974gu,PhysRevLett.33.244,Banks:1981nn} in
non-abelian gauge theories, suggesting that the current scenario could shed
some light on this related  phenomenon in quantum chromodynamics.  In the
present work, we aim at clarifying option~{\itshape ii)} on the two-loop level.

The rest of the paper is organized as follows. In the next section, we present the continuum $(2+1)$-dimensional Gross--Neveu--Yukawa field theory with two adjoint order parameters coupled to fermions,~\cite{Herbut:2022zzw} and present its analytic continuation to $d=4-\epsilon$ (see Appendix~\ref{sec:continuation} for a detailed derivation).
Then, in Section~\ref{sec:RG2loop}, we elaborate on our automated two-loop renormalization procedure and explicitly state our results for the two-loop $\beta$-functions and anomalous dimensions in $d=4-\epsilon$.
In Section~\ref{sec:fixedpointanalysis}, we perform a fixed-point analysis, both in the symmetric subspace of the model studied by quantum Monte Carlo calculations for a generalized number of fermion flavors~$N_f$, and in the full, unconstrained parameter space for $N_f = 2$, which corresponds to graphene.
Finally, in Section~\ref{sec:conclusions}, we discuss and summarize our results.

\section{Gross--Neveu--Yukawa model with two order parameters\label{sec:model}}

The model we study is given in Refs.~[\onlinecite{PhysRevLett.128.117202}] and [\onlinecite{Herbut:2022zzw}]
and describes a Gross--Neveu--Yukawa field theory in $d=2+1$ Euclidean spacetime dimensions
with an eight-component complex fermion field $\psi$ that is coupled to two
triplets of order parameters.
These order parameters are denoted in the following as
$a_i t_i$ and $b_j t_j$, with $i,j=1,2,3$ and where
$a_i$ and $b_j$ are real scalar fields with $t_{k} = \frac{\sigma_{k}}{2}$, where $\sigma_k$ denote the usual
Pauli matrices satisfying the Clifford algebra $\{\sigma_k, \sigma_l \} = 2 \delta_{k l}\mathbbm{1}$.
The scalars transform as $(1, 0) + (0, 1)$ under the global $\SO(4) \simeq \SO(3) \times \SO(3)$ symmetry group,
which is locally isomorphic to the product of two distinct $\SU(2)$ groups, i.e., $\SU(2)_{\mathrm{A}} \times \SU(2)_{\mathrm{B}}$.
\begin{table}[t]
  \begin{center}
  \caption{
      Field content of $4-\epsilon$ dimensional theory in
      Eq.~\eqref{eq:bareLagr4d} and its transformation properties under the
      global $\SU(2)_{\mathrm{A}}\times\SU(2)_{\mathrm{B}}$ symmetry and the
      Lorentz group.  $n$ labels the number of Dirac-bidoublet fermions, and we
      have suppressed their  $\SU(2)_{\mathrm{A}} \times \SU(2)_{\mathrm{B}}$
      indices for brevity.\\
    \label{table:quantities4d}}
      \begin{tabular}{|cccc|}
      \hline
       & $\SU(2)_{\mathrm{A}}$ & $\SU(2)_{\mathrm{B}}$ & Lorentz\Large\strut\\
      \hline
      $\Psi_{n}$ & $\mathbf{2}$ & $\mathbf{2}$ & $\left(\frac{1}{2}, 0\right) + \left(0, \frac{1}{2}\right)$\large\strut\\
      $a_i$ & $\mathbf{3}$ & -- & -- \large\strut\\
      $b_j$ & -- & $\mathbf{3}$ & -- \large\strut\\
     \hline
  \end{tabular}
  \end{center}
\end{table}

Using the projective $\SU(2)\simeq \SO(3)$ isomorphism
we couple the fermionic field to the
order parameters.
The corresponding $d=2+1$ Lagrangian in {\itshape Euclidean} spacetime reads
\begin{equation}\label{eq:Lagd3eight}
  \Lag_{d=2+1} =\Lag_\psi + \Lag_{ a b} + \Lag_{\psi a b}\,,
\end{equation}
with fermionic, bosonic, and Yukawa-type contributions\footnote{
Since Eq.~\eqref{eq:Leucl} is in Euclidean $2+1$ spacetime dimensions $\partial_\mu\partial^\mu=\partial_\tau^2 + \partial_x^2+\partial_y^2$.}
\begin{align}
  \Lag_\psi&=\psi^{\dagger} \partial_\tau \psi + \psi^{\dagger}\Big[\mathbbm{1}_{4 \times 4} \otimes \sum_{k=1}^{2} \sigma_k \left(-i \frac{\partial}{\partial x_k}\right)\Big] \psi \,,\\
\Lag_{a b}&=
\frac{1}{2} \big[ (\partial_\mu a_i)(\partial^\mu a_i)   +(\partial_\mu b_j)(\partial^\mu b_j) \big]\nonumber\\
    &-\frac{r_{a}}{2} (a_i a_i) \!-\! \frac{r_b}{2} (b_j b_j)  \nonumber\\
    &- \frac{\lambda_a}{8} \left(a_i a_i \right)^2
    - \frac{\lambda_b}{8} \left( b_j b_j \right)^2 - \frac{\lambda_{a b}}{4} (a_i a_i) (b_j b_j) \,,\label{eq:Leucl}\\
    \Lag_{\psi a b}\!&=\!-\psi^{\dagger} \Big[g_{a}\!\left(a_i t_i\right)\!\otimes\!\mathbbm{1}_{2\times 2}
    \!+\! g_b \mathbbm{1}_{2\times 2}\!\otimes\!\left(b_j t_j\right)\Big]\!\otimes\!\beta \psi\,,
\end{align}
where $\beta = \sigma_3$ is hermitian and anticommutes with $\sigma_1$ and $\sigma_2$.
In Ref.~[\onlinecite{Herbut:2022zzw}] the eight-component fermion field $\psi$ is split into $N_f = 2$ four-component fermion fields.
Note that compared to Ref.~[\onlinecite{Herbut:2022zzw}] we have rescaled the fields and couplings by constant factors.

The Yukawa couplings, $g_{a/b}$, and the quartic couplings, $\lambda_{a/b/ab}$, are not marginal
in $d=3$, thus, we cannot perform standard perturbation-theory computations in $d=3$.
However, if the theory admits a continuation in  $d=4-\epsilon$ dimensions, where the couplings 
are marginal, and has perturbative fixed points for small $\epsilon$, it is amenable to computations within the
framework of $\epsilon$-expansion.\cite{WILSON197475}
This can be used to provide quantitative extrapolations of fixed-point dynamics 
in the $d=3$ theory, which is the case for the model in Eq.~\eqref{eq:Lagd3eight}.

\begin{figure*}[t]
  \centering
  \includegraphics{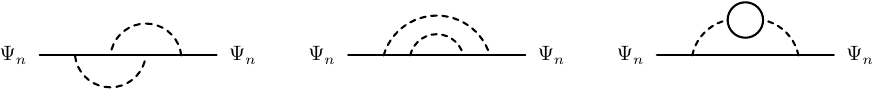}\\[1em]
  \includegraphics{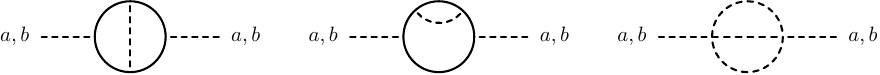}
  \caption{Example Feynman diagrams entering the computation of the 
   fermionic (upper) and bosonic (lower)  two-point functions
   at the two-loop level. Such diagrams contribute to the 
   respective field renormalization constants $Z_\Psi$ and $Z_{a/b}$. 
    \label{fig:2point}}
\end{figure*}

To this end, it is essential to formulate the theory such that it can be continuously
interpolated from $d=4$ to $d=3$ spacetime dimensions.
In Appendix~\ref{sec:continuation}, we present in detail how to continue the theory of Eq.~\eqref{eq:Lagd3eight}
in $d=4-\epsilon$. Here, we present the final result of the
continuation in Minkowskian spacetime.
It features the scalars, $a_i$ and $b_j$, whose continuation
is straightforward, and an $N_\Psi$ number of four-component Dirac fields, $\Psi_n$,
which transform as bidoublets under the global symmetry group $\SU(2)_{\mathrm{A}} \times \SU(2)_{\mathrm{B}}$.
In \autoref{table:quantities4d} we summarize the field content of the theory and its transformation
properties.

The corresponding $d=4-\epsilon$ Lagrangian in Minkowskian spacetime with $N_\Psi$ number of 
four-component Dirac-bidoublet fermions, $\Psi_n$, reads
\begin{align}\label{eq:bareLagr4d}
    \Lag =&\
    \overline{\Psi}_n i \partial_\mu \gamma^\mu \Psi_n + \frac{1}{2} \left[\left(\partial_\mu a_{i}\right)\left(\partial^\mu a_{i}\right)+\left(\partial_\mu b_{j}\right)\left(\partial^\mu b_{j}\right)\right]\nonumber\\
      &- \frac{r_a}{2} (a_i a_i) - \frac{r_b}{2} (b_j b_j) \nonumber \\
      &- \frac{\lambda_{a}}{8} \left(a_i a_i \right)^2
      - \frac{\lambda_{b}}{8} \left(b_j b_j \right)^2
      - \frac{\lambda_{ab}}{4} (a_i a_i) (b_j b_j)\nonumber  \\
      &- g_{a} \overline{\Psi}_n  \bigl(a_i t_i \bigr) \Psi_n
      - g_{b}  \overline{\Psi}_n \bigl(b_j t_j \bigr) \Psi_n \,,
\end{align}
with $\mu = 0,1,2,3$ and $\gamma^\mu$ the $4\times 4$ Dirac matrices.
The Lagrangian is manifestly Lorentz invariant.
For brevity, we have kept the contractions of the $\SU(2)_{\mathrm{A}/\mathrm{B}}$ indices implicit.

The steps required to derive the Lagrangian in Eq.~\eqref{eq:bareLagr4d}
starting from the $d=3$ Euclidean Lagrangian in Eq.~\eqref{eq:Lagd3eight}
are presented in detail in Appendix~\ref{sec:continuation}.
First, we repackage the eight-component fermion field, $\psi$, of the $d=3$ Lagrangian
into a single two-component Dirac-bidoublet fermion, $\Psi^{d=3}_{aA}$.
Next, we make Lorentz invariance manifest by re-expressing the $\beta$- and $\sigma$-matrices
in terms of Euclidean gamma matrices, and subsequently rotating to Minskowski where
$\{ \gamma^0, \gamma^1, \gamma^2 \} = \{ \beta, - \beta \sigma_1, - \beta \sigma_2\}$ 
with $\{\gamma^\mu, \gamma^\nu \} = 2 g^{\mu \nu}$ and $g^{\mu \nu} = \text{diag}(+, -, -)$.
The final step is to continue the theory from $d=3$ to $d=4$. 
In the former, Dirac fermions are complex two-component fields, whereas in the latter they are complex 
four-component fields, which implies that the na\"ive continuation of the $d=3$ Lagrangian to 
$d=4$ contains twice as many fermionic degrees of freedom.
Therefore, in order to describe graphene, i.e., a single $\Psi^{d=3}_{aA}$, from the $4-\epsilon$
Lagrangian in Eq.~\eqref{eq:bareLagr4d} within the $\epsilon$-expansion 
we must set $N_\Psi = 1/2$.
This corresponds to $N_f = 4 N_\Psi$ in the definition 
of $N_f$ from Ref.~[\onlinecite{Herbut:2022zzw}].

\section{Renormalization group at two-loop level\label{sec:RG2loop}}

We compute the $\beta$ functions and anomalous dimensions of the $d$-dimensional theory in Eq.~\eqref{eq:bareLagr4d} by the standard renormalization
procedure: fields and couplings are renormalized and the value of their
renormalization constants are extracted from the divergent terms of appropriate
Green's functions.  The renormalization constants then fix the $\beta$ functions
and anomalous dimensions of the theory.  We work within dimensional
regularization in $d=4-\epsilon$ spacetime dimensions and employ the
mass-independent $\overline{\rm MS}$ renormalization scheme.  In this section,
we introduce our notation, briefly describe the relevant computations for
determining the Renormalization Group Equations (RGEs) at the two-loop level,
and present our results.

Specifically, we renormalize the bare fields and couplings in
Eq.~\eqref{eq:bareLagr4d} (denoted with the superscript ``(0)'' below) via
\begin{align}
    \Psi_{n}^{(0)} &= Z_{\Psi}^{1/2} \Psi_{n}\,,&
     a_{i}^{(0)} &= Z_{a}^{1/2} a_{i}\,,& \nonumber\\
     b_{j}^{(0)} &= Z_{b}^{1/2} b_{j}\,,&
     r_{a,b}^{(0)} &= r_{a,b} + \delta r_{a,b}\,,& \nonumber\\
     g_{a}^{(0)} &= \mu^{\epsilon/2} Z_{g_a} g_{a}\,,&
     g_{b}^{(0)} &= \mu^{\epsilon/2} Z_{g_b} g_{b}\,,\nonumber\\
     \lambda_{a, b, ab}^{(0)} &= \mu^{\epsilon} \left(\lambda_{a,b,ab} + \delta\lambda_{a,b,ab}\right).&&&&\label{eq:bareQuantities4d}
\end{align}
Note that all fields and the Yukawa couplings, $g_{a/b}$, are multiplicatively
renormalized, while the order parameters, $r_{a,b}$, and the quartics,
$\lambda_{a,b,ab}$, must be additively renormalized. The independence of the bare
coupling on the artificial $\overline{\rm MS}$ scale $\mu$ defines the RGEs in
the standard way.

\begin{figure*}[t]
  \centering
  \includegraphics{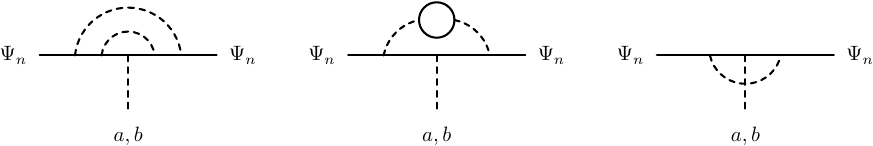}
  \caption{Example Feynman diagrams entering the computation of the
           fermion--fermion--scalar three-point functions at the two-loop level
           relevant for extracting the renormalization constants of the Yukawa couplings, $Z_{g_{a/b}}$.
    \label{fig:3point}}
\end{figure*}

\begin{figure*}[t]
  \centering
  \includegraphics{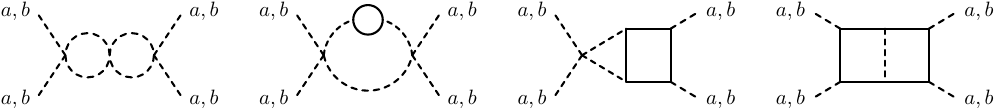}
  \caption{Example Feynman diagrams entering the computation of the
           scalar four-point functions at the two-loop level
           relevant for extracting the renormalization constants of the quartic
         couplings, $\delta\lambda_{a/b/ab}$.
    \label{fig:4point}}
\end{figure*}

We have determined all the above renormalization constants up to two-loop
accuracy by explicit computation and renormalization of offshell, one-particle-irreducible Green's functions.
The fermion and scalar field renormalization constants are extracted from the
fermion and scalar two-point functions, respectively; in \autoref{fig:2point}
we show representative two-loop diagrams for the fermion two-point function, and for the scalar two-point function.  The scalar
two-point function additionally fixes the additive renormalization constant for
the order parameters, i.e., $\delta r_{a,b}$.  To determine the renormalization of
the Yukawa couplings, $Z_{g_{a,b}}$, we renormalize the
fermion--fermion--scalar three-point functions (see \autoref{fig:3point}
for some representative two-loop diagrams) and to determine the renormalization
of quartics, $\delta \lambda_{a,b,ab}$, we renormalize the four-scalar
four-point functions (see \autoref{fig:4point}).

To extract the renormalization constants from the UV poles
of offshell, one-particle irreducible
Green's functions it suffices to compute those in terms
of an expansion in small external momenta (and masses).
The expansion leads to scaleless loop-integrals that vanish in dimensional
regularization due to the cancellation of IR and UV divergences \cite{doi:10.1063/1.1666512}.
We thus employ the method of \textit{Infrared Rearrangement}~(IRA),\cite{Chetyrkin_1998} which introduces an artificial mass $m_{\mathrm{IRA}}$ that serves as an IR
regulator. This enables us to extract the UV divergences from computing
``simple'' two-loop tadpole diagrams\footnote{In this context, we refer
  to ``tadpole'' diagrams as diagrams in which the propagators
  do not depend on the external momenta due to the expansion.}
  that depend on a single mass, i.e., $m_{\mathrm{IRA}}$.
The recursion relations and the corresponding master integrals are given
in Ref.~[\onlinecite{Bobeth:1999mk}].
Even though intermediate steps of the computation can depend
on the artificial mass $m_{\mathrm{IRA}}$
-- counterterms proportional to $m_{\mathrm{IRA}}^2$ may be required
to cancel all subdivergences -- the final renormalization constants
do not depend on $m_{\mathrm{IRA}}$. For further details, we refer to
the original discussion in Ref.~[\onlinecite{Chetyrkin_1998}].
In our case, the only such ``IRA counterterm'' that can and does enter
the computation corresponds to scalar mass terms, i.e.,
$m_{\mathrm{IRA}}^2 a_i a_i$ and $m_{\mathrm{IRA}}^2 b_j b_j$.

The two-loop calculation involves hundreds of different diagrams.
We thus make use of the automated setup \texttt{MaRTIn},\cite{MaRTIn} which employs \texttt{QGRAF}~\cite{Borowka_2017} for the diagram generation
and \texttt{FORM}~\cite{Kuipers:2012rf} routines to evaluate the amplitudes.
\texttt{MaRTIn} performs the IRA, the tensor reduction to scalar integrals,
the reduction to master integrals, and the Dirac algebra.
We performed all computations by using our own implementation of
the model in Eq.~\eqref{eq:bareLagr4d} and by extending \texttt{MaRTIn}
with a custom routine to perform the $\SU(2)_{\mathrm{A}, \mathrm{B}}$ algebra.
All one-loop results agree with the correspondig calculations in Ref.~[\onlinecite{Herbut:2022zzw}].
As an indepenent check of the explicit two-loop computation, we have
obtained the two-loop RGEs using the toolkit \texttt{ARGES}~\cite{ARGES} and
find an exact agreement.
An additional check of the computation is provided by available
two-loop $\beta$-functions of the chiral and the statistical Heisenberg model,\cite{PhysRevD.96.096010}
as these models are subsectors of our model with two triplet order parameters.

Having obtained the renormalization constants in Eq.~\eqref{eq:bareQuantities4d},
we readily obtain the $\beta$-functions and anomalous dimensions of the theory
from the independence of the bare quantities on the scale $\mu$.
The continuum RGEs given below are directly related to the ones
in the \textit{Wilsonian Renormalization Group} (WRG) by substituting
$\frac{d}{d \log \mu} \to - \frac{d}{d \log b}$, with $b$ controlling
the inclusion of IR modes in the momentum-shell integration of the WRG.
In what follows, we collect the resulting RGEs for the $d=4-\epsilon$ theory
at two-loop accuracy.
Note that in the corresponding RGEs in Ref.~[\onlinecite{Herbut:2022zzw}]
the quantity $N_f$ denotes the number of four-component fermion fields,
which is related to the number of Dirac bidoublets, $N_\Psi$, via
$N_f = 4 N_\Psi$ (cf.~\autoref{sec:model}).

\subsection*{\boldmath Yukawa $\beta$-functions}

In the $d=4-\epsilon$ theory, the $\beta$-functions for the Yukawa couplings, $g_{a,b}$,
are defined as
\begin{align}
  \beta_{g_X}(\epsilon)\equiv\frac{d g_X^2}{d\log \mu} = -\epsilon g_X^2
  +  \sum_{i=1}^\infty \Bigl(\frac{1}{16\pi^2}\Bigr)^{n} \beta_{g_X}^{(n)}\,,
\end{align}
with $X\in\{a,\,b\}$.
We find for the one- and two-loop terms
\begin{align}
  \beta_{g_a}^{(1)} =&
                      \frac{1+8 N_\Psi}{2} g_a^4
                     +\frac{9}{2} g_a^2 g_b^2
                     \,,\\
  \beta_{g_a}^{(2)} =&
                      \frac{47-192 N_\Psi}{32}g_a^6
                     -\frac{33+120 N_\Psi}{16}  g_a^4 g_b^2 \nonumber\\
                     &-\frac{177+336 N_\Psi}{32} g_a^2 g_b^4
                    +\frac{3}{2} \lambda_{ab}^2       g_a^2
                     -9 g_a^2  g_b^2\lambda_{ab} \nonumber\\
                     &-5 g_a^4       \lambda_a
                     +\frac{5}{2} g_a^2 \lambda_a^2\,,
\end{align}
and analogously $\beta_{g_b}(\epsilon) = \{\beta_{g_a}(\epsilon)~\text{with}~a\leftrightarrow b\}$.

\subsection*{\boldmath Quartic $\beta$-functions}

Similarly, the $\beta$-functions for the quartics are defined as
\begin{align}
  \beta_{\lambda_X}(\epsilon)\equiv\frac{d \lambda_X}{d\log \mu} = -\epsilon \lambda_X
  +  \sum_{i=1}^\infty \Bigl(\frac{1}{16\pi^2}\Bigr)^{n} \beta_{\lambda_X}^{(n)}\,,
\end{align}
with $X\in\{a,\,b,\,ab\}$. For $\lambda_a$ we find at one- and two-loop level
\begin{align}
  \beta_{\lambda_a}^{(1)} =&
                           -4 N_\Psi g_a^4
                           +8 N_\Psi g_a^2 \lambda_a
                           +11 \lambda_a^2
                           +3 \lambda_{ab}^2
                     \,,\\
  \beta_{\lambda_a}^{(2)} =&
                           +8  N_\Psi g_a^6
                           +24 N_\Psi g_a^4 g_b^2
                           -3  N_\Psi g_a^4 \lambda_a
                           -15 N_\Psi g_a^2 g_b^2\lambda_a\nonumber\\
                           &
                           +12 N_\Psi g_a^2 g_b^2\lambda_{ab}
                           -44 N_\Psi g_a^2 \lambda_a^2
                           -12 N_\Psi g_b^2 \lambda_{ab}^2\nonumber\\
                           &
                           -69 \lambda_a^3
                           -15 \lambda_a \lambda_{ab}^2
                           -12 \lambda_{ab}^3
                     \,,
\end{align}
and by analogy $\beta_{\lambda_b}(\epsilon) = \{\beta_{\lambda_a}(\epsilon)~\text{with}~a\leftrightarrow b\}$.
For the $\lambda_{ab}$  $\beta$-function we instead find
\begin{align}
  \beta_{\lambda_{ab}}^{(1)} =&
                             -12 N_\Psi g_a^2 g_b^2
                             +4  N_\Psi (g_a^2+g_b^2) \lambda_{ab}\nonumber\\
                             &
                             +5 (\lambda_a+\lambda_b) \lambda_{ab}
                             +4 \lambda_{ab}^2
                     \,,\\
  \beta_{\lambda_{ab}}^{(2)} =&
                             +32 N_\Psi (g_a^4 g_b^2 + g_a^2 g_b^4)
                             +10 N_\Psi g_a^2 g_b^2 (\lambda_a+\lambda_b)                             \nonumber\\&
                             +N_\Psi    g_a^2 g_b^2 \lambda_{ab}
                             -\frac{11}{2} N_\Psi (g_a^4+g_b^4) \lambda_{ab} \nonumber\\&
                             -8  N_\Psi (g_a^2+g_b^2) \lambda_{ab}^2
                             -20 N_\Psi (g_a^2 \lambda_a + g_b^2 \lambda_b)\lambda_{ab} \nonumber\\&
                             -30 (\lambda_a+\lambda_b) \lambda_{ab}^2
                             -\frac{25}{2} (\lambda_a^2+\lambda_b^2) \lambda_{ab}
                             -11 \lambda_{ab}^3\,.
\end{align}

\subsection*{Anomalous dimensions for order parameters}
The anomalous dimensions $\gamma_{r_{a,b}}$ control the RGEs for the order parameters, $r_{a,b}$, via
\begin{align}
  \gamma_{r_X} \equiv
  \frac{d \log r_{X}}{d\log\mu} = \sum_{i=1}^\infty \Bigl(\frac{1}{16\pi^2}\Bigr)^{n} \gamma_{r_X}^{(n)}\,, \label{eq:ADM_order_param}
\end{align}
where $X\in\{a,\,b\}$ and with the one- and two-loop contributions
\begin{align}
  \gamma_{r_a}^{(1)} =&4 N_\Psi g_a^2 +5 \lambda_a + 3\frac{r_b}{r_a}\lambda _{ab} \,,\\
  \gamma_{r_a}^{(2)} =&
  -\frac{N_\Psi}{2}g_a^2 \left(15 g_b^2+11 g_a^2+40 \lambda_a\right)
  -\frac{25}{2} \lambda_a^2
  -\frac{3}{2} \lambda_{ab}^2 \notag \,,\\
  & +\frac{r_b}{r_a} \biggl(N_\Psi g_b^2\left(6 g_a^2 -12 \lambda _{ab}\right)-6 \lambda _{ab}^2\biggr)\,,
\end{align}
and by analogy $\gamma_{r_b} = \{\gamma_{r_a}~\text{with}~a\leftrightarrow b\}$.
These RGEs depend on the ratio $r_b/r_a$ because the order parameters
receive additive renormalization corrections.
The sign convention in Eq.~\eqref{eq:ADM_order_param} implies that the
scaling dimension of the bosons on the fixed points, $\theta_{r_{a, b}}$,
receives the quantum correction
\begin{equation}
    \theta_{r_{a, b}} = D_{r_{a, b}} - \gamma_{r_{a, b}}^*\,,
\end{equation}
 with $D_{r_{a, b}}=2$ the engineering/classical dimension of the respective order parameter
 and $\gamma_{r_{a, b}}^*$ its anomalous dimension evaluated at the given fixed point.

\subsection*{Scalar anomalous dimensions}
Using the field renormalization constants, $Z_{a,b}$, we can also
define anomalous dimensions for the bosonic degrees of freedom $a_i$ and $b_j$ as
\begin{align}
  \gamma_{X} \equiv
  \frac{d \log Z_{X}}{d\log\mu} = \sum_{i=1}^\infty \Bigl(\frac{1}{16\pi^2}\Bigr)^{n} \gamma_{X}^{(n)}\,,
\end{align}
where $X\in\{a,\,b\}$ and with
\begin{align}
  \gamma_{a}^{(1)} =& N_\Psi g_a^2 \,,\\
  \gamma_{a}^{(2)} =&
                     \frac{5}{8}\lambda_a^2
                    +\frac{3}{8}\lambda_{ab}^2
                    -\frac{7}{8}  N_\Psi g_a^4
                    -\frac{15}{8} N_\Psi g_a^2 g_b^2\,.
\end{align}
By analogy, $\gamma_b = \{\gamma_a~\text{with}~a\leftrightarrow b\}$.

\subsection*{Fermion anomalous dimension}
For completeness, we define here in analogy to the bosonic sector
also the ``fermion anomalous dimension'' based on
the fermion-field renormalization
\begin{align}
  \gamma_\Psi \equiv
  \frac{d \log Z_{\Psi}}{d\log\mu} = \sum\nolimits_{i=1}^\infty \Bigl(\frac{1}{16\pi^2}\Bigr)^{n} \gamma_\Psi^{(n)}\,,
\end{align}
where the one- and two-loop contributions read
\begin{align}
  \gamma_\Psi^{(1)} =& \frac{3}{16} \left(g_a^2+g_b^2\right)\,,\\
  \gamma_\Psi^{(2)} =& -\frac{9}{16} N_\Psi \left(g_a^4+g_b^4\right)-\frac{9}{128}\left(g_a^2+g_b^2\right)^2\,.
\end{align}
\vspace{0.5em}

\noindent
In the next section, we use the $\beta$-functions and anomalous dimensions
of the $d=4-\epsilon$ theory presented here to explore the RG fixed-point structure
of the theory and to study the $d=3$ theory.

\section{Fixed points and critical behavior\label{sec:fixedpointanalysis}}
In this section, we analyze the fixed points of the $\beta$-functions
at the two-loop order.
First, to compare to the quantum Monte Carlo (QMC) data from Ref.~[\onlinecite{PhysRevLett.128.117202}],
we constrain the model to the higher-symmetric subspace in which
$g_a^2=g_b^2$, $\lambda_a=\lambda_b$, and $\lambda_{ab} = \lambda_{c}$.
We extend the analysis to general flavor number to explore possible fixed-point
collisions and recurrences along the $N_f = 4 N_\Psi$ axis.
Secondly, we explore the fixed-point structure and stability of the full model
containing the five independent couplings in Eq.~\eqref{eq:Lagd3eight} for the
graphene case, i.e., $N_f=2$.
We extend the leading-order analysis, which identified two stable fixed points,\cite{Herbut:2022zzw}
to the next order in the $\epsilon$-expansion.
We also present results beyond the strict fixed-order $\epsilon$-expansion
by numerically evaluating the fixed points directly from the two-loop $\beta$-functions.
In this case, we find numerous fixed-point collisions as a function of $\epsilon$.

To determine the stability of each fixed point with coordinates $c_j^{*} \in \{ g_a^{2*},
g_b^{2*}, \lambda_a^*, \lambda_b^*, \lambda_{ab}^*\}$, we linearize the RG
flow in its vicinity:
\begin{equation}
  \beta_{c_{i}} = S_{i j} \left( c_{j} - c_j^{*} \right) + \mathcal{O}\left( (c - c^{*})^2 \right),
\end{equation}
where $S_{i j}$ denotes the \textit{stability matrix},
\begin{equation}\label{eq:stability_matrix}
    S\left( \{ c_n \} \right)_{i j}
    \equiv \frac{\partial \beta_{c_i}\left( \{ c_n \} \right)}{\partial c_j} \Big|_{c^{*}} .
\end{equation}
Using the five eigenvalues of the stability matrix, we define the quantity
$\theta_k$, with $k=1,\dots,5$, as the $k$th eigenvalue, multiplied by a factor
of $-1$.  In the following, we assess the stability of a fixed point based on
the sign of the corresponding $\theta_k$ and will refer to the $\theta_k$'s as the eigenvalues of
the stability matrix.
Specifically, a positive value of $\theta_k$ corresponds
to an IR repulsive (unstable, relevant) direction, whereas a negative
value corresponds to an IR attractive (stable, irrelevant) direction of
the RG flow.
We denote a fixed point as stable when all directions in the RG flow are IR attractive.
In our analysis, the bosonic order parameters $r_a$ and $r_b$ are always set to zero,
as they correspond to relevant directions in the RG flow.

\subsection{Symmetric subspace for general $N_f$}

In the symmetric subspace relevant to the model studied by QMC, we have
\begin{align}
  &g_a^2=g_b^2=g^2 \,, \quad \lambda_a=\lambda_b=\lambda \,, \quad \lambda_{ab} = \lambda_c\,.
\end{align}
\begin{widetext}
The $\beta$-functions at two-loop order then reduce to
\begin{align}
\beta_{g^2} =-\epsilon g^2
            & + \frac{1}{16 \pi^2} g^4(5+N_f) \notag\\
           & + \frac{1}{(16 \pi^2)^2} g^2\left[ g^4 \left( -6 N_f - \frac{49}{8} \right)- g^2 \left( 5 \lambda + 9 \lambda_c \right)+ \frac{5}{2} \lambda^2 + \frac{3}{2} \lambda_c^2
  \right]\,,\\[8pt]
\beta_{\lambda}= -\epsilon\lambda
               &+ \frac{1}{16 \pi^2} \left[ -N_fg^4+ 2N_f g^2 \lambda +11 \lambda ^2+3 \lambda_c^2 \right] \notag\\
              &+ \frac{1}{(16 \pi^2)^2} \left[ 8 N_f g^6 +N_fg^4 \left(3 \lambda_c-\frac{9}{2}\lambda\right)-N_fg^2 \left(11 \lambda ^2+3 \lambda_c^2\right)-3 \left(23 \lambda ^3+5 \lambda  \lambda_c^2+4 \lambda_c^3\right) \right]\,,\\
 \beta_{\lambda_c}=-\epsilon  \lambda _c
                  &+ \frac{1}{16 \pi^2} \left[-3 N_f g^4 + 2 N_f g^2 \lambda_c +2 \lambda_c (5 \lambda +2 \lambda_c) \right]\notag\\
                 &+\frac{1}{(16 \pi^2)^2} \left[ 16 N_f g^6 +N_f g^4 \left(5 \lambda -\frac{5}{2}\lambda_c\right)-2 N_f g^2 \lambda_c (5 \lambda +2 \lambda_c)-\lambda_c (5 \lambda +\lambda_c) (5 \lambda +11 \lambda_c) \right]\,.
\end{align}
\end{widetext}
Ref.~[\onlinecite{Herbut:2022zzw}] showed that the one-loop level of these
equations feature fixed-point solutions for $N_f < N_{c}^{<} \approx
0.0164$. Moreover, the one-loop equations also admit a fixed-point solution for
$N_f > N_{c}^{>} \approx 16.83$, but not for the physically relevant case, $N_f=2$.

\begin{figure}[t]
    \centering
    \includegraphics{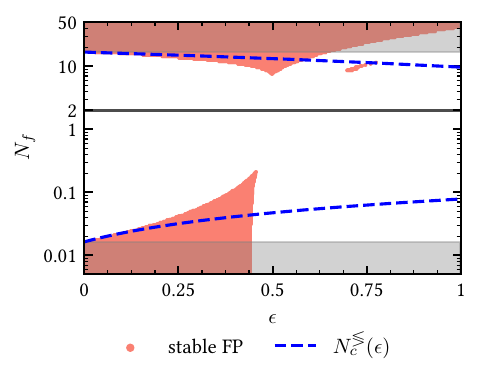}
    \caption{
     The red dots mark the positions in the $\epsilon-N_f$ plane where a
     fixed-point (FP) solution at two-loop order exists.
     The gray shaded areas indicate the region in which a FP exists at one-loop order.
     The solid, black horizontal line shows the case $N_f=2$ and
     the dashed, blue lines correspond to the extrapolation of the two-loop correction in
     Eqs.~\eqref{eq:Nf_c_gt} and \eqref{eq:Nf_c_lt} for small $\epsilon$.
     \label{fig:symNf}}
\end{figure}

To search for possible fixed points in the two-loop $\beta$-functions, we first
evaluate the full set of equations for fixed $\epsilon$ and for various values
of $N_f$.
This approach enables us to
find new fixed points only present at the two-loop order. Hence, it
differs from the strict fixed-order $\epsilon$-expansion, where we compute
perturbative corrections to one-loop fixed points.
In \autoref{fig:symNf}, we have summarized our findings. Concretely, we mark positions in the $\epsilon-N_f$ plane
with a red dot, where a stable fixed-point solution of the two-loop $\beta$-functions exists.
In addition, the gray regions show values of $N_f$, where the leading-order (LO) $\beta$-functions indicate the existence of a stable fixed-point solution.
Moreover, we display next-to-leading order (NLO) corrections to this critical number of fermions as blue dashed lines, which we will
derive below in Eqs.~\eqref{eq:Nf_c_gt} and \eqref{eq:Nf_c_lt}.
Notably, our numerical approach not only exposes outliers among
stable solutions, e.g., $\left( \epsilon, N_f\right) \sim (0.75, 10)$, but also exhibits
sharp kinks in the boundary of the stable fixed-point solutions,
which would not appear in the strict $\epsilon$-expansion.
Very similar behavior can already be observed within the well-known purely bosonic case, $g=0$. For example, for the decoupled fixed point (DFP) where
$\lambda^*_c = 0$, the explicit solution of the
NLO $\beta$-functions reads $\lambda^* = (11+ \sqrt{121-276\epsilon})/552$. This becomes complex for $\epsilon \geq 121/276 \approx 0.438$ and hence does not correspond to a physically admissible solution anymore.
As expected, our findings agree with the LO results of Ref.~[\onlinecite{Herbut:2022zzw}] for
small $\epsilon$, i.e., the critical values of $N_f$ coincide in the limit $\epsilon \rightarrow 0$, both
at LO and NLO (see dashed-dotted lines in \autoref{fig:symNf}).

We already know that to first order in $\epsilon$, the case $N_f=2$ does not
feature a fixed point, i.e., we cannot build on that at the next order in
$\epsilon$.  However, using the full numerical solution, we can track whether
the two-loop corrections bring the values for $N_{c}^{\lessgtr}$ closer to
$N_f=2$, by setting up the corresponding series in $\epsilon$, i.e.,
\begin{align}\label{eq:Nf_c_gt}
    N_{c}^{>} \approx 16.83 + \delta N_{c}^{>\,(\text{2-loop})} \epsilon +\mathcal{O}(\epsilon^2)\,,
\end{align}
where $\delta N_{c}^{> \,(\text{2-loop})}$ is to be determined, see Refs.~[\onlinecite{PhysRevB.100.134507}]~and~[\onlinecite{PhysRevLett.78.980}].
Our setup for the numerical approximation of
the slope uses a step size in $\epsilon$ of width $\Delta_\epsilon = 0.002$ and
in the flavor number, we use 200 evenly-spaced logarithmic values in the ranges $1/2 \leq N_f \leq 50$.
We start at
$\epsilon_{\mathrm{start}} = 0.002$ and determine the boundary separating
stable and unstable solutions, and calculate the slope for small $\epsilon$
including points up to $\epsilon = 0.15$ using linear regression, while fixing
the intercept at the one-loop result $N_{c}^{>}(\epsilon=0) = 16.83$.  We
accordingly extract $\delta N_{c}^{> \,(\text{2-loop})}$ from \autoref{fig:symNf} as
\begin{align}
    \delta N_{c}^{> \, (\text{2-loop})}\approx - 7.14\,,
\end{align}
and analogously, using a different range for the flavor number, i.e., $0.0001 \leq N_f \leq 0.5$,
determine the lower boundary
\begin{align}\label{eq:Nf_c_lt}
    N_{c}^{<} \approx 0.0164 + \delta N_{c}^{< \,(\text{2-loop})}\epsilon +\mathcal{O}(\epsilon^2)\,,
\end{align}
with
\begin{align}
\delta N_{c}^{< \,(\text{2-loop})} \approx 0.062\,.
\end{align}
The two lines corresponding to Eqs.~\eqref{eq:Nf_c_gt} and \eqref{eq:Nf_c_lt}
are depicted by the blue, dashed lines in \autoref{fig:symNf}.

We conclude that the two-loop order brings the critical flavor number significantly 
closer to the physical case of $N_f=2$. Therefore, it may still be possible that
the true critical flavor number $N_{f,c}$ is close or even below $N_f=2$,
which would allow critical, or at least pseudo-critical scaling as we discuss
further below.

\subsection{Unconstrained parameter space for $N_f=2$\label{subsec:eps_exp_2_graphene}}

Building upon the results of Ref.~[\onlinecite{Herbut:2022zzw}] and the two-loop results presented
in \autoref{sec:RG2loop}, we first evaluate the perturbative fixed points within the $\epsilon$-expansion
at two-loop accuracy. Here, we restrict the analysis to the case corresponding to graphene, and
thus fix $N_f = 2$. At one-loop level, the $\beta$-functions for the
Yukawa and the quartic sector decouple, allowing for an independent study of
the two subsectors. This is no longer the case at the two-loop order.
The values of the couplings at the fixed points admit a perturbative expansion for small $\epsilon$:
\begin{align}
    c^{*}_i = D_i \epsilon + F_i \epsilon^2 + \mathcal{O}(\epsilon^3),
\end{align}
with $c_i \in \{ g_a^2,\,g_b^2,\, \lambda_a,\, \lambda_b,\, \lambda_{ab}\}$.
The one-loop $\beta$-functions determine the coefficients $D_i$, whereas we compute
$F_i$ by solving the system of two-loop $\beta$-functions:
\begin{align}
  \forall i : \beta_{c_i}\left( \{ c^{*}_j \} \right) = 0,
\end{align}
where $\beta_{c_i}$ is the $\beta$-function of the coupling $c_i$.

To compactly present the fixed points, we label the fixed-point coordinates
in the Yukawa sector as $Y_j$ and in the bosonic one as $B_k$.
The tuple formed by the two coordinates describes the fixed point: $(Y_j,B_k)$.
We express the fixed-point couplings in terms of the rescaled couplings
\begin{equation*}
  \bar{g}_{a,b}^2 \equiv \frac{g_{a,b}^{* \,2}}{16\pi^2}\,,\qquad
  \bar{\lambda}_{a,b,ab} \equiv \frac{\lambda^{*}_{a,b,ab}}{16\pi^2}\,,
\end{equation*}
introduced for brevity.

We will not discuss the already extensively studied family of purely bosonic
fixed points, i.e.,
\begin{equation}
  Y_1 \, : \, \bar{g}_a^2 = 0 \,,\,\,\bar{g}_b^2 = 0\,,
\end{equation}
in which available calculations have identified stable fixed-point solutions at a precision beyond NLO.\cite{PhysRevB.67.054505}
Concretely, for $N_f=2$ we obtain a total of four different perturbative fixed 
points with physical values for all couplings at $\mathcal{O}(\epsilon^2)$:
\begin{equation}
  \label{eq:pert_fp_2L}
  \begin{split}
        (Y_2, B_{1b}),\qquad&    (Y_2, B_{2b}),\\
        (Y_3, B_{1a}),\qquad&    (Y_3, B_{2a}).
  \end{split}
\end{equation}
\begin{figure*}[t]
     \centering
     \includegraphics{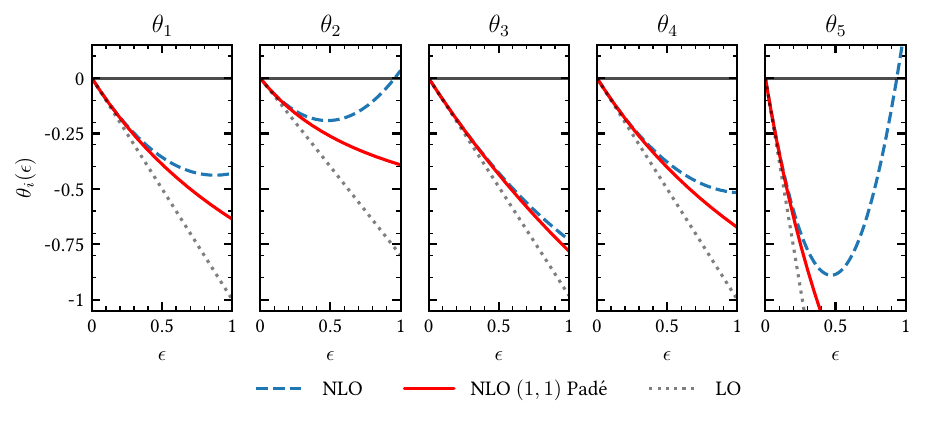}
     \caption{
       The five eigenvalues of the stability matrix in Eq.~\eqref{eq:stability_matrix}
       multiplied by a factor of $-1$, i.e., $\theta_k$, as a function of $\epsilon$,
       evaluated at the decoupled fixed points
       $(Y_2, B_{1b})$ or equivalently $(Y_3, B_{1a})$.
       The gray dotted, blue dashed, and solid red lines correspond to 
       the LO, NLO, and NLO $(m=1,n=1)$-Padé results, respectively.
       Taking the limit $\epsilon \rightarrow 1$, two of the $\theta_k$'s become positive at
       NLO, making the fixed point unstable (IR repulsive). Considering,
       however, the NLO Padé approximant, the fixed points retain their stability 
       at two-loop order. 
 \label{fig:stab_matrix_ev_Y2B1b}}
\end{figure*}
In terms of the rescaled couplings, the coordinates of the fixed points in Eq.~\eqref{eq:pert_fp_2L}
in the Yukawa sector read
\begin{align}
  Y_2 \,    : \bar{g}_a^2 &= 0,\quad \bar{g}_b^2 = \frac{2}{5} \epsilon + \frac{11689}{60500} \epsilon ^2 ,&&\\
  Y_3 \,    : \phantom{\bar{g}_b^2} &\!\!\{ Y_{2}\quad\text{with} \quad a\leftrightarrow b\},&
\end{align}
and in the bosonic sector we have $\bar{\lambda}_{ab} = 0$ and
\begin{align}
  B_{1a} \, :      \bar{\lambda}_a &= \frac{8}{55} \epsilon\! +\! \frac{311246}{3161125} \epsilon^2 ,
                  &\bar{\lambda}_b &= \frac{1}{11} \epsilon\! +\! \frac{69 }{1331} \epsilon^2 ,\\
  B_{2a} \, :      \bar{\lambda}_a &= \frac{8}{55} \epsilon\! +\! \frac{311246}{3161125} \epsilon^2 ,
                  &\bar{\lambda}_b &= 0,\\[0.5em]
  B_{1b} \, :     \phantom{\bar\lambda_a}& \!\!\{ B_{1a}\quad\text{with} \quad a\leftrightarrow b\},& && \\[0.5em]
  B_{2b} \, :     \phantom{\bar\lambda_a}& \!\!\{ B_{2a}\quad\text{with} \quad a\leftrightarrow b\}.& && 
\end{align}

We find by explicit computation 
that further fixed-point solutions associated to the Yukawa 
sector $Y_4\, : \bar{g}_a^2 = \bar{g}_b^2 \neq 0$
have complex solutions for the 
quartic couplings: $\bar{\lambda}_{a/b/ab} \notin \mathbb{R}$.
Since these fixed points do not constitute physical solutions, we do not include them
in the following discussion. We note, however, that the existence of physical
and stable solutions does depend on the number of fermions, $N_f$. This is
apparent also from our previous discussion of the symmetric subsystem in which 
we extrapolated the critical number of fermions in the 3d case to $N_{f,c}^{<}(\epsilon = 1)
\approx 0.0784$ and $N_{f,c}^{>}(\epsilon = 1) \approx 9.69$, cf., Eqs.~\eqref{eq:Nf_c_gt} and \eqref{eq:Nf_c_lt}.

The one-loop fixed-point analysis on Ref.~[\onlinecite{Herbut:2022zzw}]
showed that the two fixed points
\begin{equation}\label{eq:1L_stable_fps}
  (Y_2, B_{1b}) \,,\quad(Y_3, B_{1a})\,,
\end{equation}
were the only stable ones in the physical 3d case, obtained by taking the limit
$\epsilon \rightarrow 1$.
As all fixed points of the system, they are symmetric with respect 
to exchanging the labels $a \leftrightarrow b$ in all the couplings.
Given that the mixed quartic coupling, $\lambda_{ab}$,
is zero at these fixed points, they are commonly denoted as DFPs.

When including the two-loop corrections, we see that up to second order in the $\epsilon$-expansion, 
two of the five eigenvalues of the stability matrix become positive
close to $d=3$, see \autoref{fig:stab_matrix_ev_Y2B1b}, making the one-loop-stable fixed points in
Eq.~\eqref{eq:1L_stable_fps} unstable. Considering, however, the $(m=1,n=1)$
\textit{Padé approximant} (the $m=0, n=2$ one is divergent), all five
eigenvalues stay negative, indicating that the two fixed points retain their
one-loop stability even in the physical case of $\epsilon \rightarrow 1$.
In Appendix~\ref{app:ev_eps_full_system} we provide the full expression for the five
eigenvalues of the stability matrix for all fixed points in Eq.~\eqref{eq:pert_fp_2L}
up to $\mathcal{O}(\epsilon^2)$. 
We thus find that in the 3d limit there are no further stable fixed points 
beyond the one-loop ones after including the two-loop corrections.
In \autoref{fig:stab_matrix_ev_Y2B1b}, we show the five eigenvalues of the two candidate
stable fixed points as a function of $\epsilon$ at LO, NLO, and NLO $(1,1)$-Padé.
We see that all eigenvalues are negative as $\epsilon\to 1$ at NLO $(1,1)$-Padé (red solid lines).

A further check of our computation is provided by
Aharony's exact scaling relation\cite{PhysRevLett.88.059703}
\begin{equation}\label{eq:aharony}
    \theta_{a b} = \frac{1}{\nu_a} + \frac{1}{\nu_b} - d\,,
\end{equation}
which relates the critical exponents $\nu_a$ of the 3d Heisenberg universality
class and $\nu_b$ of the chiral Heisenberg universality class,\cite{Herbut:2022zzw,PhysRevD.96.096010}
to the scaling dimension $\theta_{a b}$ of the mixed quartic coupling
$\lambda_{a b}$, evaluated at the DFPs in Eq.~\eqref{eq:1L_stable_fps}.
Using the anomalous dimensions of the order parameters $a_i t_i$ and $b_j t_j$,
as given in Eq.~\eqref{eq:ADM_order_param}, we can determine the critical
exponents $\nu_{a, b}$ at the DFPs. Our conventions imply that
\begin{equation}
  \frac{1}{\nu_{a, b}} = D_{r_{a, b}} - \gamma_{r_{a, b}}^*\,,
\end{equation}
with $D_{r_a}=2$ the engineering/classical dimension of the respective order
parameter and $\gamma_{r_{a, b}}^*$ the value of the anomalous dimension
evaluated on the (decoupled) fixed point. The one-loop result for the critical
exponents of the (chiral) Heisenberg universality class are in agreement with
the discussion in Ref.~[\onlinecite{Herbut:2022zzw}]. Including the two-loop
contribution we find
\begin{equation}
    \frac{1}{\nu_a} = 2 - \frac{5}{11} \epsilon - \frac{415}{2662}\epsilon^2 + {\cal O}(\epsilon^3)\,,
\end{equation}
and analogously for the critical exponent of the chiral Heisenberg universality
class:
\begin{equation}
  \frac{1}{\nu_b} = 2 - \frac{84}{55} \epsilon + \frac{2576729}{6322250}\epsilon^2 + {\cal O}(\epsilon^3)\,,
\end{equation}
which are in agreement with the literature.\cite{herbut2007modern,PhysRevD.96.096010,PhysRevB.107.035151}
Using Aharony's scaling relation in Eq.~\eqref{eq:aharony} and $d=4-\epsilon$, we find
\begin{equation}
    \theta_3 = \theta_{a b} = - \frac{54}{55} \epsilon + \frac{795552}{3161125} \epsilon^2 + {\cal O}(\epsilon^3)\,,
\end{equation}
which corresponds to the NLO eigenvalue $\theta_3$ 
(cf.~\autoref{fig:stab_matrix_ev_Y2B1b} and Appendix~\ref{app:ev_eps_full_system}).
The scaling dimension $\theta_{a b}$ becomes negative in the limit
$\epsilon \rightarrow 1$, which indicates that the coupling $\lambda_{a b}$ is
irrelevant at two-loop order, thus supporting the
stability of the Heisenberg+chiral Heisenberg fixed point, as indicated by beyond NLO calculations, cf. Ref.~\onlinecite{Herbut:2022zzw}.

We conclude that the perturbative $\epsilon$-expansion ansatz does
not unambiguously settle the stability of the fixed points in the
physically relevant 3d case.
Specifically, expanding the eigenvalues of all the perturbative fixed points in
Eq.~\eqref{eq:pert_fp_2L} at NLO shows that the one-loop-stable fixed points
in Eq.~\eqref{eq:1L_stable_fps} lose their stability at $\epsilon \approx
0.95$ and that the system exhibits no stable perturbative fixed point
in the 3d case after taking $\epsilon \rightarrow 1$.
Making use, however, of the $(1,1)$ NLO Padé approximant, the two fixed points do
retain their one-loop stability.

\begin{figure}[t]
    \centering
    \includegraphics[width=0.95\columnwidth]{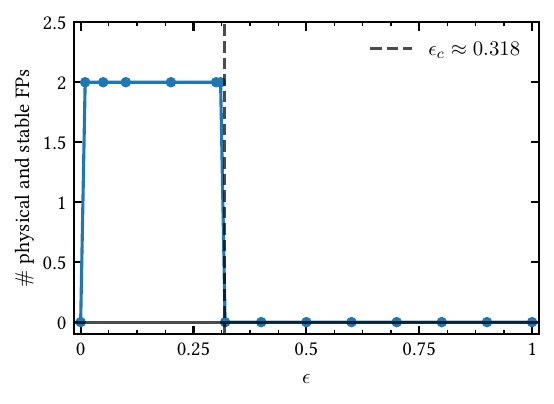}
    \caption{
      The number of stable fixed points (FPs) with a physical value of
      all couplings, i.e., $\forall i : c^*_i \geq 0$, as a function of
      $\epsilon$, found by the numerical solution of the corresponding $\beta$-functions.
      For small $\epsilon$, the $\epsilon$-expansion predicts two 
      stable FPs, cf., Eq.~\eqref{eq:1L_stable_fps}. 
      For $\epsilon = 0$, the system exhibits no stable FP, 
      as both perturbative FPs coincide with the unstable Gaussian FP.  
      The blue dots mark the values of $\epsilon$, for which we performed the 
      numerical study.
      At a critical value of $\epsilon$, i.e., $\epsilon_c \approx 0.318$,
      we observe that the two stable FPs lose their stability by
      colliding with new FPs that are only present at two-loop order.
      After crossing the critical threshold $\epsilon_c$, the system exhibits no 
      stable, physical FPs, in particular not in the 3d case, 
      obtained by taking the limit $\epsilon \rightarrow 1$.
      This result retains higher-order terms in $\epsilon$ and appears 
      to be in tension with the previous findings using the strict
      $\epsilon$-expansion predictions (cf.~\autoref{fig:stab_matrix_ev_Y2B1b}), whose NLO
      $(1,1)$ Padé approximant predicted the existence of two physical and stable 3d FPs.
  \label{fig:Num_phys_fp}}
\end{figure}

We close this section with a brief numerical study of the fixed-point
solutions of the $\beta$-functions that goes beyond the strict $\epsilon\to0$
limit.
To this end, we vary $\epsilon \in [0, 1.0]$; for small values of $\epsilon$
we expect good agreement with the perturbative expansion.
A thorough analysis shows that,
starting from $\epsilon > 0$, the perturbative, stable fixed-point candidates in
Eq.~\eqref{eq:1L_stable_fps} remain stable, until they collide with two new
non-perturbative and unstable fixed points only present at two-loops at a
critical $\epsilon_c \approx 0.318$. After crossing this critical threshold, no
stable fixed point exists, in particular not in the 3d case.  For $\epsilon = 0$,
the system exhibits no stable fixed point, as both perturbative fixed points
coincide with the unstable Gaussian one.  We have summarized our
non-perturbative findings in \autoref{fig:Num_phys_fp}, where we show the
dependence of the amount of stable fixed points with physical values for all
the couplings, i.e., $\forall i : c_i \geq 0$, against $\epsilon$.

These findings are thus in agreement with the NLO prediction from the
two-loop $\beta$-functions using the strict $\epsilon$-expansion
but are in tension with the corresponding NLO $(1,1)$-Padé (see \autoref{fig:stab_matrix_ev_Y2B1b}),
which states that the one-loop-stable fixed points in Eq.~\eqref{eq:1L_stable_fps} retain their stability at two-loops even in the physical 3d case, i.e., taking the limit $\epsilon \rightarrow 1$.
\section{Discussion\label{sec:conclusions}}

In summary, we have investigated a Gross--Neveu--Yukawa field theory describing 
a number of $N_f$ Dirac fermions coupled to two triplets of $\SO(3)$ order
parameters, employing the perturbative renormalization group at two-loop level
in $4-\epsilon$ dimensions. The study was motivated by the recent numerical
observation of relativistic quantum criticality, employing large-scale quantum
Monte Carlo simulations.\cite{PhysRevLett.128.117202} 
This criticality was found to be in conflict with the absence of a stable
renormalization-group fixed point of the above Gross--Neveu--Yukawa theory for
the physically relevant number of Dirac fermions, i.e.,  $N_f=2$.  In fact, at
the one-loop level a stable fixed point only appears in the field theory for
large, $N_f > N_{c}^{>} \approx 16.8$ or very small $N_f < N_{c}^{<}\approx
0.0164$. For the intermediate range $N_f \in [N_{c}^{<},N_{c}^{>}]$, the 
fixed point disappears into the complex plane due to a fixed-point collision 
and annihilation.~\cite{PhysRevLett.124.161601,PhysRevD.98.096014,PhysRevLett.32.292,PhysRevLett.44.837,PhysRevLett.76.4588,10.1143/PTP.105.809,PhysRevB.71.184519,Gies2006,herbut2007modern,PhysRevD.80.125005,PhysRevLett.113.106401,PhysRevD.94.025036,PhysRevB.95.075101,PhysRevB.100.134507,gorbenkoWalkingI2018,Gorbenko_2018}  
In the present work, we investigated both the symmetric 
subspace studied by quantum Monte Carlo simulations and the full, 
unconstrained parameter space of the model. In the unconstrained parameter 
space, we found a pair of stable fixed points at NLO $(1,1)$-Padé, which are, 
however, not part of the symmetric subsystem.

For the symmetric subsystem, a key result of our present renormalization-group study at two-loop order
is the $\epsilon$ correction to the upper and lower critical flavor numbers, i.e., $N_{c}^{\lessgtr}$.
Both corrections drive the critical flavor numbers closer to
the physical case $N_f=2$ with respect to the one-loop result.
However, the magnitude of the corrections as $\epsilon\to 1$ is not 
large enough to make the fixed points stable for the physical system, which thus
still suggests the absence of criticality in $2+1$ dimensions.

Hence, our two-loop study cannot fully resolve the conflicting findings between
the quantum Monte Carlo data and the field-theoretical approach.  Still, the
two-loop corrections do point towards the possibility that the
renormalization-group fixed point might be restored at even higher-loop orders,
e.g., by further suppressing the value of $N_{c}^{>}$ towards $N_f=2$, similar
to what is observed in the abelian Higgs model at four
loops.\cite{PhysRevB.100.134507}

We note that even the case of, e.g., $N_{c}^{>} \gtrsim 2$, would be very
interesting, because it could naturally explain the observation of approximate
power laws and a large correlation length, reminiscent of critical behavior,
despite the absence of true criticality. Such behavior could occur due to
pseudo-critical walking and Miransky scaling, appearing when the
renormalization-group flow is governed by an underlying complex conformal field
theory (CFT).\cite{kaplanConformality2009,gorbenkoWalkingI2018}

A complex CFT is the natural consequence of a fixed-point collision and
annihilation as observed in our study and is expected to be a common phenomenon
in many physical systems.\cite{kaplanConformality2009}  However, a numerical
confirmation is inherently difficult due to the similarity of continuous and
very weak first-order behavior in a finite-size system.  A distinct detection
tool for future quantum Monte Carlo studies could be the analysis of the
entanglement entropy as, e.g., pushed forward recently in the context of
deconfined quantum phase transitions.\cite{PhysRevLett.128.010601,Song:2023wlg}

For future field-theoretical studies, it could be very interesting to extract
even higher-order corrections to the critical number of fermion flavors in
order to consolidate or refute such complex CFT scenario.

\subsubsection*{Acknowledgments}
We thank L.~Di~Pietro for stimulating discussions.  MMS acknowledges funding
from the Deutsche Forschungsgemeinschaft (DFG, German Research Foundation)
within Project-ID 277146847, SFB 1238 (project C02), and the DFG Heisenberg
programme (Project-ID 452976698).
MU acknowledges support by the Erasmus+ programme of the European Union and is grateful to, both the University of Trieste and the ICTP for the kind hospitality during the preparation of this work. IFH is supported by the NSERC of Canada.

\appendix

\section{Continuation from $d=3$ to $d=4$\label{sec:continuation}}

In this section, we continue the Euclidean $d=3$
Lagrangian in Eq.~\eqref{eq:Lagd3eight} to $d=4-\epsilon$ dimensions. For convenience, we also
make the transition to Minkowskian spacetime since all our computations are
performed in it. We start in $d=3$-dimensional Euclidean spacetime, where
Dirac fermions are described by complex two-component spinors and furthermore
add a label ``$\mathrm{E}$'' to indicate when the objects are in Euclidean
spacetime. 
Since bosons continue trivially into higher dimensions, we focus
on the fermion sectors, i.e., $\Lag^\mathrm{E}_\psi$ and $\Lag^\mathrm{E}_{\psi ab}$.

The tensor product in the Lagrangian in Eq.~\eqref{eq:Lagd3eight}
\begin{equation}
    M_1 \otimes M_2 \otimes M_3 \,,
\end{equation}
with $M_1 , M_2 , M_3$ being $2 \times 2$ matrices, 
acts upon the two valley, two sublattice, and two spin degrees of freedom of the eight-component
fermion field $\psi$, cf., Ref.~[\onlinecite{Herbut:2022zzw}].
We rewrite this tensor notation by rearranging the eight components of $\psi$
into the two components of a complex Dirac fermion in $d=3$ that is a bidoublet 
of $\SU(2)_{\mathrm{A}}\times \SU(2)_{\mathrm{B}}$:
\begin{equation}
  \label{eq:psiPsi}
  \psi \longrightarrow \Psi_{\alpha,aA}^{d=3} \,.
\end{equation}
Here, Greek letters ($\alpha=1,2$) denote the spin components, which 
will end up being the Lorentz spinor indices, 
lowercase Latin letters ($a = 1, 2$) are the fundamental indices of 
$\SU(2)_{\mathrm{A}}$ (valley degrees of freedom), and uppercase 
Latin letters ($A=1,2$) are the fundamental indices of 
$\SU(2)_{\mathrm{B}}$ (sublattice degrees of freedom).
Lowered $\SU(2)_{\mathrm{A}/\mathrm{B}}$ indices are always assumed to 
be in the fundamental representation of $\SU(2)$ (as in Eq.~\eqref{eq:psiPsi}),
whereas raised ones denote indices in the anti-fundamental representation.
By contracting with $\epsilon^{ab}$ or $\epsilon^{AB}$ we can raise (or lower)
$\SU(2)$ indices to relate the fundamental and anti-fundamental representations.

We reserve the letters $i,j = 1,2,3$ to be used as adjoint indices of
$\SU(2)_{\mathrm{A}/\mathrm{B}}$. The order parameters transform in 
the adjoint representation of their respective $\SU(2)$, which
can be obtained from  the traceless part of the $\mathbf{2}\otimes \mathbf{\bar 2}$
tensor product. Therefore, to construct invariants, it is useful to express
the order parameters as
\begin{equation}
  (a_i t_i)_a^{\;\; b} \,,\quad(b_j t_j)_A^{\;\; B} \,,
\end{equation}
with $t_i=\sigma_i/2$.

Finally, we define the conjugate field $\Psi^{\dagger} \equiv \overline{\Psi} \beta \equiv \overline{\Psi} \gamma^0_{\mathrm{E}}$
with $\beta^2 = \sigma_3^2 = \mathbbm{1}_2$ to find by rewriting Eq.~\eqref{eq:Lagd3eight}:
\begin{align}\label{eq:appLag3}
  \Lag^{\mathrm{E}}_\Psi         &= \overline{\Psi}^{d=3}_{aA}\big[\slashed \partial_\mathrm{E} \epsilon^{ab}\epsilon^{AB}\big]\Psi^{d=3}_{bB}\,, \\ 
  \Lag^{\mathrm{E}}_{\Psi a b} &= -\overline{\Psi}^{d=3}_{aA} \big[
                                                               g_a \left(a_i t_i \right)^{a b} \epsilon^{A B} 
                                                             + g_b \epsilon^{a b}\left(b_j t_j\right)^{A B}   \big] \Psi^{d=3}_{bB}\,,\nonumber 
\end{align} 
where we have suppressed the spinor indices. Here,
we identified the $d=3$ gamma matrices in Euclidean spacetime
as $\{ \gamma_\mathrm{E}^0, \gamma_\mathrm{E}^1, \gamma_\mathrm{E}^2 \} \equiv \{ \beta, -i \beta \sigma_1, -i \beta \sigma_2
\} $, which satisfy the Clifford algebra 
$\{ \gamma^\mu_\mathrm{E}, \gamma^\nu_\mathrm{E} \} = 2 \delta^{\mu \nu}$.

The transition from Euclidean to Minkowskian spacetime is thereafter
straightforward by performing a Wick rotation, i.e., rotating the Euclidean
time, $\tau$, to the Minkowskian time, $t$: $t \rightarrow -i \tau$.  
This rotation implies replacing $\slashed \partial_{\mathrm{E}} \rightarrow
i \slashed \partial \equiv i\partial_\mu \gamma^\mu$, where 
$\gamma^\mu$ denotes the gamma matrices in Minkowskian spacetime,
which satisfy $\{ \gamma^\mu, \gamma^\nu \} = 2 g^{\mu \nu}$ with $g^{\mu \nu}
= \text{diag}(+, -, -)$. Concretely, the gamma matrices in Minkowskian spacetime 
are related to their Euclidean counterparts via 
$\gamma^0 \rightarrow \gamma^0_{\mathrm{E}} = \beta$ and 
$\gamma^k \rightarrow -i \gamma^k_{\mathrm{E}} = -\beta \sigma^k$ for $k=1,2$.  
Thus, we rewrote the original Euclidean $d=3$ Lagrangian with an eight-component complex fermion
field $\psi$ into a Lagrangian in Minkowskian spacetime, featuring a single
two-component complex Dirac fermion that transforms as a bidoublet
under the global symmetry group $\SU(2)_{\mathrm{A}}\times \SU(2)_{\mathrm{B}}$.\\

Next we derive our continuation of the $d=3$ Lagrangian to 
$d=4-\epsilon$ dimensions. Our Dirac fermions in $d=3$ are 
two-component complex spinors, whereas $d=4$ Dirac fermions are complex, four-component 
spinors or equivalently two, complex two-component Weyl spinors. Therefore, na\"ively
we expect that the continuation will double the fermionic degrees of freedom. We shall
see that this is indeed the case if we couple the fermions to adjoint order parameters.

The attempt to derive a continuation without doubling  the fermionic degrees of freedom
by interpreting the $d=3$ two-component Dirac field, $\Psi^{d=3}_{aA}$,
as a corresponding two-component Weyl field in $d=4$, $\xi_{aA}$, is not possible.
The reason is twofold: 
{\itshape i)} for $\xi_{aA}$ there is no Lorentz-invariant counterpart to the $d=3$ Yukawa
term, e.g., $\overline{\Psi}^{d=3} (a_i t_i)\Psi^{d=3}$, 
since in four Minkowskian spacetime dimensions, the representation of the Lorentz group
is complex, which is not the case in $d=3$. 
Thus, while  complex conjugation in $d=3$ gives rise to an equivalent representation---allowing for 
Dirac mass terms---it leads to a nonequivalent representation in $d=4$, i.e., there is no notion
of chirality in $d=3$.
{\itshape ii)} there is a Lorentz-variant way to couple $\xi_{aA}$ with
adjoint order parameters. However, this contraction vanishes due to the global
symmetry structure:
\begin{equation}\label{eq:yuk_4d_zero}
  \xi_{aA}  (a_i t_i)^{ab} \epsilon^{AB} \xi_{bB} = 0\,,
\end{equation}
(analogously for $(b_j t_j)^{A B}$),
which evaluates to zero due to the fact that $(a_i t_i)^{ab}$ is symmetric 
in the indices $a$ and $b$ and $\xi_{aA}$ is an anticommuting field.

Importantly, the analogous contraction with two independent Weyl fields, 
$\xi_{aA}$ and $\eta_{aA}$ {\itshape does not vanish}, e.g., 
\begin{equation}\label{eq:4d_weyls_yuk}
  \xi_{aA} (a_i t_i)^{a b} \epsilon^{A B} \eta_{bB}  \neq 0 \,,
\end{equation}
and analogously for $(b_j t_j)^{A B}$.
Thus, to construct a non-vanishing Yukawa interaction using Weyl fields
in $d=4$ we have to \textit{double} the fermionic degrees of freedom with 
respect to $d=3$, i.e., we need two \textit{different flavors} of Weyl fields.

In fact, the form of Eq.~\eqref{eq:4d_weyls_yuk} is precisely the one we obtain 
when we decompose the reducible, complex Dirac field, $\Psi^{d=3}_{aA}$ in
its irreducible, real two-component Majorana fields:
\begin{equation}\label{eq:majorana_decomp}
  \Psi^{d=3}_{a A} = \chi_{1,a A} + i\,\chi_{2,aA}\,.
\end{equation}
In terms of $\chi_{1}$ and $\chi_2$, the Yukawa terms read
\begin{align}
  \overline{\Psi}^{d=3}_{a A} (a_i t_i)^{a b} \epsilon^{A B} \Psi^{d=3}_{b B} 
  &= i \overline{\chi }_{1,a A} (a_i t_i)^{a b} \epsilon^{A B} \chi_{2,bB} \nonumber\\
  &- i \overline{\chi }_{2,a A} \underbrace{(a_i t_i)^{a b} \epsilon^{A B}}_{=-(a_i t_i)^{b a} \epsilon^{B A}} \chi_{1,bB} \nonumber\\
  &= 2i \overline{\chi}_{1,a A} (a_i t_i)^{a b} \epsilon^{A B} \chi_{2,bB} \neq 0 \,,
\end{align}
and analogously for $(b_j t_j)^{A B}$. This illustrates why 
doubling the fermionic degrees appears to be the natural way
to continue the theory via Eq.~\eqref{eq:4d_weyls_yuk}.

\begin{table}[t]
    \begin{center}
      \caption{
        Field content of $4-\epsilon$ dimensional theory in
        Eq.~\eqref{eq:AppBareLagr4d} and its transformation properties under the
        global $\SU(2)_{\mathrm{A}}\times\SU(2)_{\mathrm{B}}$ symmetry and the
        Lorentz group. $n$ labels the number of Dirac-bidoublet fermions.\\
      \label{table:AppQuantities4d}}
    \begin{tabular}{|cccc|}
    \hline
     & $\SU(2)_{\mathrm{A}}$ & $\SU(2)_{\mathrm{B}}$ & Lorentz\Large\strut\\
    \hline
    $\eta_{n, aA}$, $\xi_{n,aA}$ & ${\mathbf{2}}$ & ${\mathbf{2}}$ & $(1/2, 0)$ \large\strut\\
    $\Psi_{n, a A}$ & $\mathbf{2}$ & $\mathbf{2}$ & $(1/2, 0) + (0, 1/2)$ \large\strut\\
    $a_i$ & $\mathbf{3}$ & -- & -- \large\strut\\
    $b_j$ & -- & $\mathbf{3}$ & -- \large\strut\\
   \hline
\end{tabular}
\end{center}
\end{table}

Going back to the $d=4$ description we can bundle together the two different flavors
of Weyl fields, $\xi_{aA}$ and $\eta_{aA}$---taken without loss of generality to transform
under the left-chiral representation $(1/2,0)$  of Lorentz---into one four-component Dirac field,
$\Psi_{a A}$, as
\begin{align}
    \Psi_{a A} \equiv\left(\begin{array}{c}
    \eta_{\alpha, a A} \\
    \xi_{a A}^{\dagger \dot{\alpha}}
    \end{array}\right)\,,\qquad
\overline{\Psi}^{a A}=\left(\xi^{\alpha, a A}, \eta_{\dot{\alpha}}^{\dagger a A}\right)\,.
\end{align}
Here, $\alpha$ denotes the spinor index for left-chiral spinors and 
$\dot{\alpha}$ denotes the right-chiral $(0, 1/2)$ one.

The final step to fix the continuation is to ensure that we describe the
correct number of fermionic degrees of freedom, when we extrapolate
the $d=4-\epsilon$ results to $d=3$. This is not possible with a single $\Psi_{aA}$.
We achieve this instead by introducing $N_\Psi$ number of Dirac-bidoublet fermions, $\Psi_{n,aA}$.
To describe the original eight independent fermionic components of graphene in $d=3$, i.e., 
$N_f = 2$ four-component fermion fields corresponding to a single $\Psi^{d=3}_{aA}$, 
within the $\epsilon$-expansion, we must set $N_\Psi = 1/2$.
For the definition of $N_f$ from Ref.~[\onlinecite{Herbut:2022zzw}], this corresponds to $ N_f = 4 N_\Psi$.
In terms of $\Psi_{n,aA}$ the resulting, manifestly Lorentz-invariant Lagrangian in 
$d=4-\epsilon$ reads:
\begin{equation}
    \begin{split}
    \Lag &=
      \overline{\Psi}_{n} i \slashed{\partial} \Psi_{n} + \frac{1}{2} \left[\left(\partial_\mu a_{i}\right)\left(\partial^\mu a_{i}\right)+\left(\partial_\mu b_{j}\right)\left(\partial^\mu b_{j}\right)\right] \\
      &- \frac{r_a}{2} (a_i a_i) - \frac{r_b}{2} (b_j b_j) \\
      &- \frac{\lambda_{a}}{8} \left(a_i a_i \right)^2
      - \frac{\lambda_{b}}{8} \left(b_j b_j \right)^2
      - \frac{\lambda_{ab}}{4} (a_i a_i) (b_j b_j) \\
      &- g_{a} \overline{\Psi}_{n}  \bigl(a_i t_i \bigr) \Psi_{n}
      - g_{b} \overline{\Psi}_{n} \bigl(b_j t_j \bigr) \Psi_{n} \,,\\
    \end{split}
  \label{eq:AppBareLagr4d}
  \end{equation}
where for brevity we have kept all Lorentz and $\SU(2)_{\mathrm{A}/\mathrm{B}}$ contractions implicit.
In \autoref{table:AppQuantities4d} we summarize the transformation properties of all fields under 
the Lorentz group and the global symmetry $\SU(2)_{\mathrm{A}}\times\SU(2)_\mathrm{B}$.

Notice that the structure of the $d=4$ Lagrangian in terms of four-component Dirac fermions, Eq.~\eqref{eq:AppBareLagr4d},
is in total analogy to the Minkowskian version of the $d=3$ Lagrangian written in terms of 
two-component Dirac fermions, Eq.~\eqref{eq:appLag3}.

\section{Eigenvalues of the stability matrix within the $\epsilon$-expansion\label{app:ev_eps_full_system}}

In this section, we provide the explicit results for all the eigenvalues of
the stability matrix in Eq.~\eqref{eq:stability_matrix} multiplied by 
a factor of $-1$ and denoted by $\theta_k$ 
for the four perturbative fixed points of Eq.~\eqref{eq:pert_fp_2L} at NLO.

As discussed in \autoref{subsec:eps_exp_2_graphene}, taking the limit of
$\epsilon \rightarrow 1$ to obtain the 3d case yields no stable fixed-point
solutions, which corresponds to having at least one positive eigenvalue
(IR repulsive).

\begin{widetext}
Explicitly, the eigenvalues $\{ \theta_1, \theta_2, \theta_3, \theta_4, \theta_5 \}$ up to $\mathcal{O}(\epsilon^2)$ in $d=4-\epsilon$ read:
\begin{align}
    &(Y_2, B_{1b})~\&~(Y_3, B_{1a}): \, \left\{\epsilon -\frac{69 \epsilon ^2}{121},\frac{4 \epsilon }{5}-\frac{1148 \epsilon ^2}{1375},\frac{54 \epsilon }{55}-\frac{795552 \epsilon ^2}{3161125},\epsilon -\frac{11689 \epsilon ^2}{24200},\frac{19 \epsilon }{5}-\frac{1165427 \epsilon ^2}{287375}\right\} \,, \\
    &(Y_2, B_{2b})~\&~(Y_3, B_{2a}): \,  \left\{-\epsilon ,\frac{4 \epsilon }{5}-\frac{25881 \epsilon ^2}{30250},\frac{29 \epsilon }{55}-\frac{2576729 \epsilon ^2}{6322250},\epsilon -\frac{11689 \epsilon ^2}{24200},\frac{19 \epsilon }{5}-\frac{1165427 \epsilon ^2}{287375}\right\} \,.
\end{align}
\end{widetext}

Using the NLO eigenvalues, we can furthermore calculate $(m,n)$ Padé approximants
\begin{equation}
    \theta(\epsilon) = \frac{\sum_{j=0}^m D_j \epsilon^j}{1+\sum_{k=0}^n F_k \epsilon^k} \,.
\end{equation}
Since the $m=0, n=2$ Padé approximant diverges, the only non-trivial one to consider is the $m=n=1$ approximant.
As a result, we find that only for the two fixed points $(Y_2, B_{1b})$ and $(Y_3, B_{1a})$ all eigenvalues stay negative (IR attractive), 
in particular in the $(2+1)$-dimensional case, indicating that these fixed points retain their LO stability.

\addcontentsline{toc}{section}{References}
\bibliography{references}

\end{document}